\documentclass[12pt]{article}

\usepackage[a4paper, margin=2cm]{geometry}

\usepackage{setspace}
\usepackage[utf8]{inputenc}
\usepackage[T1]{fontenc}
\usepackage[english]{babel}

\usepackage[super,numbers]{natbib} 

\usepackage{hyperref}
\usepackage{graphicx}
\usepackage{float}
\floatstyle{plaintop}
\restylefloat{table}
\usepackage{placeins}
\usepackage{subfig}
\usepackage{gensymb}
\usepackage{orcidlink}
\usepackage{graphicx,amsmath,bm,amssymb,xcolor,here,soul}
\usepackage{psfrag,multirow}
\usepackage{url}
\usepackage{siunitx}
\usepackage[T1]{fontenc} 
\usepackage{stackengine}
\usepackage{float}
\usepackage{xcolor}
\usepackage{mathtools}
\usepackage[utf8]{inputenc}
\usepackage{textgreek}
\DeclareUnicodeCharacter{2666}{\ensuremath{\diamondsuit}} 
\usepackage{hyperref} 
\usepackage{orcidlink}

\title{Grain-boundary-mediated kinetic arrest in graphite-to-diamond transformation}
\author{Zuzanna Malinowska-Trzmielak\textsuperscript{1}*, Vilmos Neuman\textsuperscript{2}, Mark Wilson\textsuperscript{3}}
\date{\today}
\begin{document}
\maketitle

\textsuperscript{1} Department of Chemistry, University of Oxford; zuzanna.trzmielak@chem.ox.ac.uk

\textsuperscript{2} Department of Chemistry, University of Oxford; vilmos.neuman@st-hildas.ox.ac.uk

\textsuperscript{3} Department of Chemistry, University of Oxford; mark.wilson@chem.ox.ac.uk

*Correspondence: zuzanna.trzmielak@chem.ox.ac.uk, Physical and Theoretical Chemistry Laboratory, Department of Chemistry, University of Oxford, OX1 3QZ

\maketitle

\begin{abstract}
 The graphite-to-diamond transition exhibits striking variability under high-pressure, high-temperature (HPHT) conditions, producing diamond, graphitic phases, or metastable, mixed diamond–graphite nanocomposites despite similar synthesis conditions. Existing atomistic models, largely based on idealised single-crystal graphite, do not explain the persistence of partially transformed intermediate states under HPHT conditions. Here, using large-scale molecular dynamics simulations, we show that precursor grain structure governs graphite-to-diamond transformation pathways by decoupling diamond nucleation from cooperative transformation propagation. Grain boundaries first facilitate local sp$^3$ nucleation, after which diamond growth propagates within individual grains but becomes arrested at crystallographically mismatched grain boundaries. As a result, structurally heterogeneous graphite stabilizes kinetically arrested mixed sp$^2$/sp$^3$ states, whereas large or single-crystalline domains favour cooperative bulk transformation into diamond. Our findings identify structural heterogeneity as a missing control parameter alongside pressure and temperature, reframing metastable transformation products as kinetically trapped states arising from precursor microstructure rather than thermodynamic intermediates. Precursor crystallinity therefore emerges as a practical control parameter governing graphite-to-diamond transformation pathways.
\end{abstract}

\section{Main}
The graphite-to-diamond transition is one of the most extensively studied solid-state transformations. Yet, under nearly identical high-pressure, high-temperature (HPHT) conditions, graphite has been observed to transform into diamond, remain graphitic, or form metastable diamond–graphite nanocomposites.\cite{nemeth_diamond-graphene_2020, luo_coherent_2022, li_discovery_2024, ohfuji_natural_2015, irifune_formation_2004} Previous computational studies have provided valuable insight into atomistic transformation pathways, but have largely focused on idealised single-crystal graphite and static transition pathways.\cite{luo_coherent_2022,fahy_theoretical_1987, scandalo_pressure_1995, khaliullin_nucleation_2011} Such approaches cannot fully account for the structural heterogeneity and finite-temperature dynamics characteristic of experimentally relevant HPHT systems. In idealised transformation pathways, the activation barrier associated with the initial formation of sp$^3$-bonded intermediate structures exceeds that of subsequent diamond growth.\cite{luo_coherent_2022} Consequently, under HPHT conditions, overcoming the initial nucleation barrier should promote continued transformation toward diamond rather than stabilisation of partially transformed intermediate states.

One possible origin of this discrepancy lies in the structural diversity of graphitic precursors. Reports of diaphite formation employ polycrystalline graphite or graphite powders as starting materials, whereas experiments using single-crystalline graphite under comparable HPHT conditions predominantly yield diamond or polycrystalline diamond.\cite{nemeth_diamond-graphene_2020,luo_coherent_2022,ohfuji_natural_2015} These observations suggest that grain boundaries and orientational disorder play a decisive role in determining transformation pathways.

\begin{figure}[!ht]
    \centering
    \includegraphics[width=16 cm]{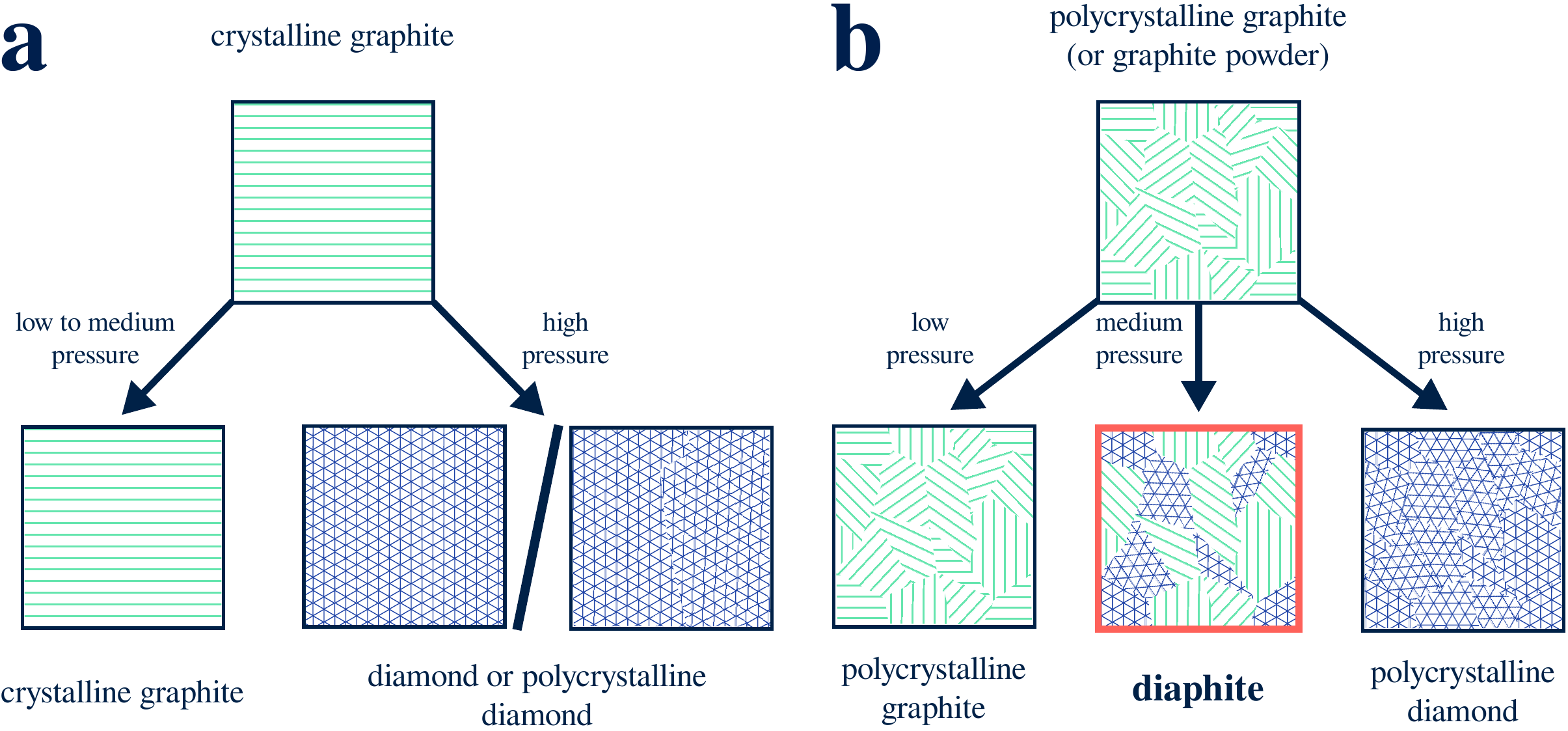}
    \caption{\textbf{Precursor structure governs graphite-to-diamond transformation pathways under HPHT conditions.} Under high-pressure, high-temperature (HPHT) conditions, single-crystalline graphite either remains graphitic or transforms cooperatively into diamond (a). In contrast, polycrystalline graphite and graphite powders exhibit spatially heterogeneous transformation pathways that can yield kinetically arrested mixed sp$^2$/sp$^3$ structures, including diaphites (b). This behaviour arises from the dual role of grain boundaries, which facilitate local sp$^3$ nucleation while frustrating cooperative diamond propagation across neighbouring graphitic domains. Green striped regions denote graphitic domains; navy blue regions denote diamond.}
    \label{fig:single_crys_poly_cryst}
\end{figure}

Here, using large-scale molecular dynamics (MD) simulations, we show that transformation outcome depends strongly on the initial graphite structure (Fig.~\ref{fig:single_crys_poly_cryst}). Ideal single-crystal graphite either remains structurally stable or transforms cooperatively into diamond, whereas polycrystalline graphite exhibits spatially heterogeneous and kinetically arrested transformation pathways, yielding either diamond–graphite nanocomposites or polycrystalline diamond. We identify grain boundaries as a dual mechanistic element that both promotes diamond nucleation and kinetically frustrates cooperative transformation propagation through crystallographic mismatch between neighbouring graphitic domains.

\begin{figure}[!t]
    \centering
    \includegraphics[width=17 cm]{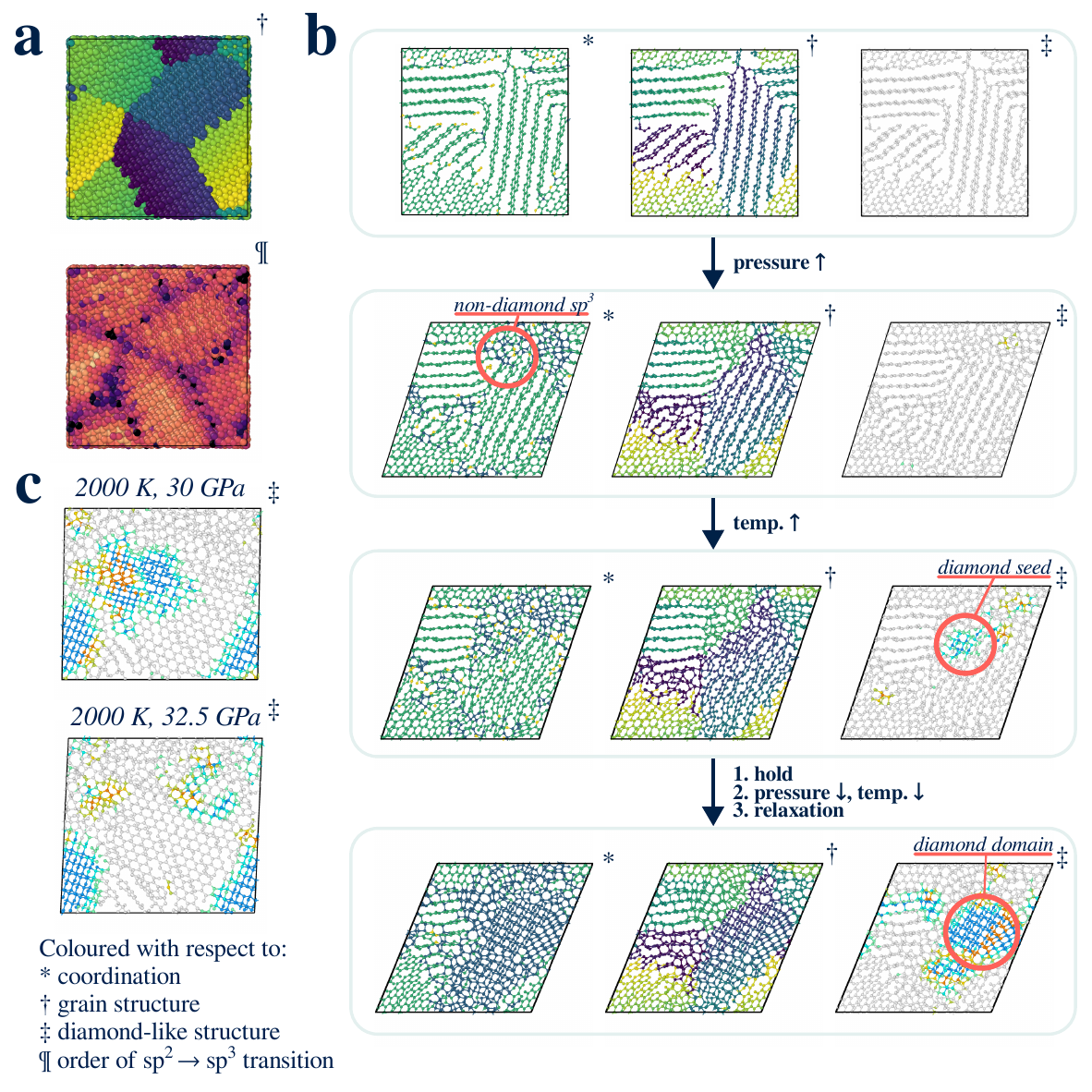}
    \caption{\textbf{Grain boundaries decouple diamond nucleation from transformation propagation.} (a) One face of a simulation cell containing ten graphitic grains. Top: grains coloured by identity. Bottom: the same face coloured by the earliest time at which atoms become four-coordinated across all 20 simulations. Darker colours indicate earlier conversion, demonstrating preferential sp$^3$ nucleation at grain boundaries. (b) Time evolution of a structural slice during HPHT synthesis. Pressurisation induces non-diamond sp\textsuperscript{3} carbon at grain boundaries, while subsequent heating enables a subset of these regions to reorganise into diamond seeds that propagate within individual grains. Propagation is arrested at grain boundaries, confining diamond domains to single grains. (c) Diamond-like sp\textsuperscript{3} content within the same structural slice as a function of applied pressure and temperature. Increasing pressure alone does not systematically increase diamond content, indicating that diamond growth is controlled by thermally activated seed formation rather than pressure-driven propagation. Where atoms are coloured by coordination, green denotes three-coordinated (graphitic) and blue denotes four-coordinated carbon.}
    \label{fig:mechanism}
\end{figure}

The mechanism explaining this variability arises from localised strain and structural disorder at grain boundaries. Upon pressurisation, four-coordinated (sp\textsuperscript{3}) carbon atoms first emerge at grain boundaries (Fig. \ref{fig:mechanism}a). At this stage, these sp\textsuperscript{3} configurations rarely adopt a diamond-like arrangement. Increasing temperature enhances atomic mobility and because grain boundaries are intrinsically less stable than the bulk, these strained regions preferentially undergo atomic rearrangement. This allows non-diamond-like sp\textsuperscript{3} regions to reorganise into the thermodynamically stable diamond motif, forming diamond seeds. (Fig. \ref{fig:mechanism}b) This rearrangement is driven by thermally activated stochastic atomic motion; consequently, increasing pressure alone does not necessarily promote continued diamond formation, as illustrated in Fig.~\ref{fig:mechanism}c. Once a diamond seed forms, this motif can propagate into the adjacent graphitic grain. The phase transition is arrested at the grain boundary, where crystallographic mismatch inhibits. Grain structure therefore governs both diamond nucleation and the extent of transformation propagation under HPHT conditions, enabling kinetically arrested mixed sp$^2$/sp$^3$ states in which only a fraction of graphitic grains transform into diamond.

\begin{figure}[!t]
    \centering
    \includegraphics[width=16.5cm]{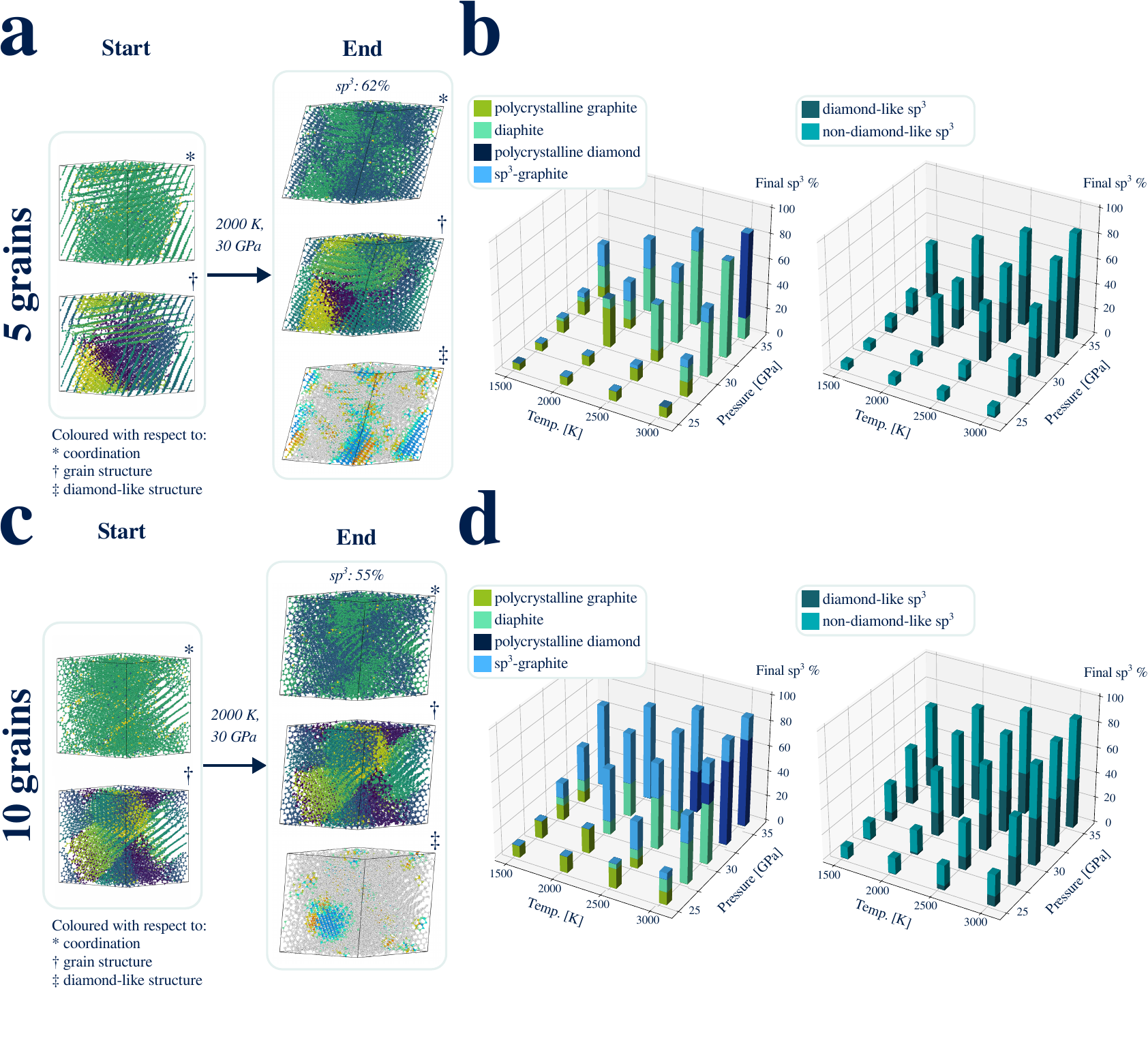}
    \caption{\textbf{Grain-boundary density governs the balance between diamond nucleation and propagation.} Comparison of systems with intermediate (5 grains) and small (10 grains) grain sizes under identical high-pressure, high-temperature conditions. (a,c) Initial and final atomic structures of 5-grain (a) and 10-grain (c) systems subjected to 2000~K and 30~GPa in molecular dynamics simulations. (b,d) Final sp$^3$ content as a function of applied pressure and temperature for the corresponding systems. The relative heights of the coloured bar segments indicate the probability of each transformation outcome. Systems with intermediate grain sizes exhibit more effective cooperative diamond propagation, producing larger diamond-like domains while retaining a clear spatial correlation with the initial grain structure. In contrast, the higher grain-boundary density in the 10-grain systems promotes the formation of non-diamond sp$^3$ carbon that frustrates propagation and yields more spatially diffuse transformation pathways. Where atoms are coloured by coordination, green denotes three-coordinated (graphitic) and blue denotes four-coordinated carbon.}
    \label{fig:grain_difference}
\end{figure}

Consistent with this mechanism, grain size exerts a non-monotonic influence on graphite-to-diamond transition. If the grains are too small, the high density of grain boundaries generates an excess of non-diamond sp\textsuperscript{3} atoms that obstruct diamond-seed propagation. As the temperature increases, diamond seeds can form, but their growth is restricted by surrounding sp\textsuperscript{3} regions with incompatible orientations. As a result, transformation remains spatially heterogeneous even at elevated temperatures and pressures, retaining a substantial fraction of non-diamond sp$^3$ carbon (Fig. \ref{fig:grain_difference}c, d).

At intermediate grain sizes, the balance between sp\textsuperscript{3} generation and diamond propagation is optimised. The reduced number of strained grain boundaries requires higher pressures to generate a sufficient population of non-diamond four-coordinated atoms (Fig. \ref{fig:grain_difference}b), while remaining low enough to permit diamond-domain propagation once a seed forms. In this regime, strain induced by pressurisation is relieved predominantly through propagative formation of the high-pressure phase, namely diamond domains. By contrast, smaller grains favour stepwise sp\textsuperscript{2}-to-sp\textsuperscript{3} conversion initiated at the grain boundaries. Consequently, intermediate-grain systems retain a strong correlation between the initial grain structure and the final phase distribution, whereas smaller-grain systems exhibit more diffuse, non-diamond-like boundaries (Fig. \ref{fig:grain_difference}a, c).

In contrast, large grains approach the behaviour of single-crystal graphite (Fig.~\ref{fig:single_crys_poly_cryst}). The reduced density of strained grain-boundary regions suppresses the formation and rearrangement of non-diamond sp$^3$ configurations into stable diamond nuclei. As a result, substantially higher pressures are required to initiate transformation within the graphitic domains themselves rather than at grain boundaries. Under these conditions, transformation increasingly proceeds through cooperative bulk propagation rather than kinetically arrested boundary-mediated pathways, yielding diamond or polycrystalline diamond.

This work establishes precursor grain structure as a decisive factor governing graphite-to-diamond transformation pathways under HPHT conditions. We show that grain boundaries simultaneously promote diamond nucleation and kinetically frustrate cooperative transformation propagation, thereby stabilising partially transformed sp$^2$/sp$^3$ states. This mechanism reframes metastable diamond–graphite nanocomposites not as arrested thermodynamic intermediates accessible through finely tuned pressure–temperature conditions, but as kinetically trapped products arising from structural heterogeneity in the graphitic precursor. Further aspects of this phase transition, in particular the structure of the developing phase, are discussed elsewhere.\cite{neuman_fractality} Although other forms of structural disorder may also contribute, our results identify grain-boundary-mediated heterogeneity as a dominant factor governing transformation pathways under HPHT conditions. In practical terms, transformation pathways should correlate with precursor crystallinity, providing a route toward rational control of graphite-to-diamond transformation behaviour. More broadly, these findings show how structural heterogeneity can decouple nucleation from transformation propagation and provide insight into kinetically arrested reconstructive phase transformations.

\section{Methods}

\subsection{Generation and selection of polycrystalline graphite structures}
An ensemble of 50 polycrystalline graphite configurations was generated by Voronoi tessellation \cite{brostow_construction_1978,finney_procedure_1979} as implemented in the \texttt{ATOMSK}\cite{hirel_atomsk_2015} software, with randomly placed seeds defining grains in a cubic cell (side length $45~\text{\AA}$) under periodic boundary conditions. To select structurally diverse samples, we computed Wasserstein distances between persistence diagrams for all configurations using \texttt{HomCloud}.\cite{obayashi_persistent_2022} We retained the ten most distinct polycrystals (five with five grains and five with ten grains), each containing $\sim 10^{4}$ atoms (10{,}016--10{,}166).

\subsection{Molecular dynamics simulations}
Molecular dynamics (MD) simulations were performed in \texttt{LAMMPS}\cite{thompson_lammps_2022} using an atomic cluster expansion\cite{drautz_atomic_2019} (ACE) potential for elemental carbon \cite{qamar_atomic_2023} which has been benchmarked against \textit{ab initio} simulations to accurately reproduces graphite, diamond, and metastable mixed sp$^2$/sp$^3$ carbon structures.\cite{malinowska-trzmielak_transferability_2026} Each structure was first quenched to the nearest local minimum by conjugate-gradient minimisation, converged to changes in energy $<10^{-12}$~eV or a global force norm $<10^{-12}$~eV\,\AA$^{-1}$. All MD simulations employed periodic boundary conditions, an integration time step of 1~fs, and the isothermal-isobaric ensemble with Nos\'e--Hoover thermostat and barostat damping constants of 1~ps and 10~ps, respectively. A triclinic barostat was used to allow relaxation of all six cell degrees of freedom.

Each trajectory followed the following protocol: equilibration at p=1~bar and T=300~K for 20~ps; pressurisation to the target pressure over 50~ps; heating to the target temperature over 50~ps; holding at target conditions for 50~ps; and then depressurisation and cooling to ambient conditions over 100~ps, followed by a final 20~ps equilibration at 1~bar and 300~K. We simulated all combinations of $T \in \{1500, 2000, 2500, 3000, 3500\}$~K and $p \in \{25, 27.5, 30, 32.5, 35\}$~GPa, for a total of 200 simulations.

\subsection{Structural analysis}
Structural analysis was performed in \texttt{OVITO}.\cite{stukowski_visualization_2009} Bonds were defined using a cut-off distance of $1.7~\text{\AA}$. Diamond-like domains were identified by common neighbour analysis \cite{maras_global_2016} as implemented in \texttt{OVITO}. Final structures were classified based on the spatial distribution of graphitic, diamond-like, and non-diamond four-coordinated domains identified through common neighbour analysis. Each structure fall into one of those categories: polycrystalline diamond (no graphitic domains), polycrystalline graphite (no diamond-like domains), diaphite (coexisting diamond and graphite domains), or sp\textsuperscript{3}--graphite composite (graphite and non-diamond four coordinated domains).

\section{Funding}
Funding from the EPSRC Centre for Doctoral Training in Inorganic Chemistry for Future Manufacturing (OxICFM), EP/S023828/1.

\section{Author contributions}
Zuzanna Malinowska-Trzmielak: conceptualization, methodology, formal analysis, investigation, writing - original draft, visualization; Vilmos Neuman: methodology, formal analysis, investigation, writing - original draft; Mark Wilson: conceptualization, resources, writing - review \& editing, supervision, project administration, funding acquisition.

\section{Competing interests}
The authors declare no competing interests.

\section{Materials \& Correspondence}
Correspondence and material requests should be addressed to Zuzanna Malinowska-Trzmielak (zuzanna.trzmielak@chem.ox.ac.uk).

\bibliographystyle{unsrtnat}
\bibliography{Polycrystallinity_paper}

@article{nemeth_diamond-graphene_2020,
	title = {Diamond-{Graphene} {Composite} {Nanostructures}},
	volume = {20},
	issn = {1530-6984},
	url = {https://doi.org/10.1021/acs.nanolett.0c00556},
	doi = {10.1021/acs.nanolett.0c00556},
	number = {5},
	urldate = {2024-09-17},
	journal = {Nano Lett.},
	publisher = {American Chemical Society},
	author = {Németh, Péter and McColl, Kit and Smith, Rachael L. and Murri, Mara and Garvie, Laurence A. J. and Alvaro, Matteo and Pécz, Béla and Jones, Adrian P. and Corà, Furio and Salzmann, Christoph G. and McMillan, Paul F.},
	month = may,
	year = {2020},
	pages = {3611--3619},
}

@article{li_discovery_2024,
	title = {Discovery of {Gradia} between {Graphite} and {Diamond}},
	volume = {5},
	url = {https://doi.org/10.1021/accountsmr.4c00029},
	doi = {10.1021/accountsmr.4c00029},
	number = {5},
	urldate = {2024-11-13},
	journal = {Acc. Mater. Res.},
	publisher = {American Chemical Society},
	author = {Li, Baozhong and Liu, Bing and Luo, Kun and Tong, Ke and Zhao, Zhisheng and Tian, Yongjun},
	month = may,
	year = {2024},
	pages = {614--624},
}

@article{luo_coherent_2022,
	title = {Coherent interfaces govern direct transformation from graphite to diamond},
	volume = {607},
	copyright = {2022 The Author(s)},
	issn = {1476-4687},
	url = {https://www.nature.com/articles/s41586-022-04863-2},
	doi = {10.1038/s41586-022-04863-2},
	language = {en},
	number = {7919},
	urldate = {2024-11-13},
	journal = {Nature},
	publisher = {Nature Publishing Group},
	author = {Luo, Kun and Liu, Bing and Hu, Wentao and Dong, Xiao and Wang, Yanbin and Huang, Quan and Gao, Yufei and Sun, Lei and Zhao, Zhisheng and Wu, Yingju and Zhang, Yang and Ma, Mengdong and Zhou, Xiang-Feng and He, Julong and Yu, Dongli and Liu, Zhongyuan and Xu, Bo and Tian, Yongjun},
	month = jul,
	year = {2022},
	pages = {486--491},
}

@article{irifune_formation_2004,
	series = {New {Developments} in {High}-{Pressure} {Mineral} {Physics} and {Applications} to the {Earth}'s {Interior}},
	title = {Formation of pure polycrystalline diamond by direct conversion of graphite at high pressure and high temperature},
	volume = {143-144},
	issn = {0031-9201},
	url = {https://www.sciencedirect.com/science/article/pii/S0031920104000780},
	doi = {10.1016/j.pepi.2003.06.004},
	urldate = {2025-12-01},
	journal = {Phys. Earth Planet. Inter.},
	author = {Irifune, Tetsuo and Kurio, Ayako and Sakamoto, Shizue and Inoue, Toru and Sumiya, Hitoshi and Funakoshi, Ken-ichi},
	month = jun,
	year = {2004},
	pages = {593--600},
}

@article{ohfuji_natural_2015,
	title = {Natural occurrence of pure nano-polycrystalline diamond from impact crater},
	volume = {5},
	copyright = {2015 The Author(s)},
	issn = {2045-2322},
	url = {https://www.nature.com/articles/srep14702},
	doi = {10.1038/srep14702},
	language = {en},
	number = {1},
	urldate = {2025-12-01},
	journal = {Sci Rep},
	publisher = {Nature Publishing Group},
	author = {Ohfuji, Hiroaki and Irifune, Tetsuo and Litasov, Konstantin D. and Yamashita, Tomoharu and Isobe, Futoshi and Afanasiev, Valentin P. and Pokhilenko, Nikolai P.},
	month = oct,
	year = {2015},
	pages = {14702},
}

@article{malinowska-trzmielak_transferability_2026, title={Transferability of carbon potentials for novel carbon polymorphs}, volume={34}, url = {https://iopscience.iop.org/article/10.1088/1361-651X/ae3e04},DOI={10.1088/1361-651x/ae3e04}, number={2}, journal={Modelling and Simulation in Materials Science and Engineering}, author={Malinowska-Trzmielak, Zuzanna and Grobert, Nicole and Wilson, Mark}, year={2026}, month={Feb}, pages={025004}}

@article{hirel_atomsk_2015,
	title = {Atomsk: {A} tool for manipulating and converting atomic data files},
	volume = {197},
	issn = {0010-4655},
	shorttitle = {Atomsk},
	url = {https://www.sciencedirect.com/science/article/pii/S0010465515002817},
	doi = {10.1016/j.cpc.2015.07.012},
	urldate = {2026-01-27},
	journal = {Comput. Phys. Commun.},
	author = {Hirel, Pierre},
	month = dec,
	year = {2015},
	pages = {212--219},
}

@article{brostow_construction_1978,
	title = {Construction of {Voronoi} polyhedra},
	volume = {29},
	issn = {0021-9991},
	url = {https://www.sciencedirect.com/science/article/pii/0021999178901109},
	doi = {10.1016/0021-9991(78)90110-9},
	number = {1},
	urldate = {2026-01-27},
	journal = {J. Comput. Phys.},
	author = {Brostow, Witold and Dussault, Jean-Pierre and Fox, Bennett L},
	month = oct,
	year = {1978},
	pages = {81--92},
}

@article{finney_procedure_1979,
	title = {A procedure for the construction of {Voronoi} polyhedra},
	volume = {32},
	issn = {0021-9991},
	url = {https://www.sciencedirect.com/science/article/pii/0021999179901463},
	doi = {10.1016/0021-9991(79)90146-3},
	number = {1},
	urldate = {2026-01-27},
	journal = {J. Comput. Phys.},
	author = {Finney, J. L},
	month = jul,
	year = {1979},
	pages = {137--143},
}

@article{qamar_atomic_2023,
	title = {Atomic {Cluster} {Expansion} for {Quantum}-{Accurate} {Large}-{Scale} {Simulations} of {Carbon}},
	volume = {19},
	issn = {1549-9618},
	url = {https://doi.org/10.1021/acs.jctc.2c01149},
	doi = {10.1021/acs.jctc.2c01149},
	number = {15},
	urldate = {2026-01-27},
	journal = {J. Chem. Theory Comput.},
	publisher = {American Chemical Society},
	author = {Qamar, Minaam and Mrovec, Matous and Lysogorskiy, Yury and Bochkarev, Anton and Drautz, Ralf},
	month = aug,
	year = {2023},
	pages = {5151--5167},
}

@article{drautz_atomic_2019,
	title = {Atomic cluster expansion for accurate and transferable interatomic potentials},
	volume = {99},
	url = {https://link.aps.org/doi/10.1103/PhysRevB.99.014104},
	doi = {10.1103/PhysRevB.99.014104},
	number = {1},
	urldate = {2026-01-27},
	journal = {Phys. Rev. B},
	publisher = {American Physical Society},
	author = {Drautz, Ralf},
	month = jan,
	year = {2019},
	pages = {014104},
}

@article{thompson_lammps_2022,
	title = {{LAMMPS} - a flexible simulation tool for particle-based materials modeling at the atomic, meso, and continuum scales},
	volume = {271},
	issn = {0010-4655},
	url = {https://www.sciencedirect.com/science/article/pii/S0010465521002836},
	doi = {10.1016/j.cpc.2021.108171},
	urldate = {2026-01-27},
	journal = {Comput. Phys. Commun.},
	author = {Thompson, Aidan P. and Aktulga, H. Metin and Berger, Richard and Bolintineanu, Dan S. and Brown, W. Michael and Crozier, Paul S. and in 't Veld, Pieter J. and Kohlmeyer, Axel and Moore, Stan G. and Nguyen, Trung Dac and Shan, Ray and Stevens, Mark J. and Tranchida, Julien and Trott, Christian and Plimpton, Steven J.},
	month = feb,
	year = {2022},
	pages = {108171},
}

@article{stukowski_visualization_2009,
	title = {Visualization and analysis of atomistic simulation data with {OVITO}–the {Open} {Visualization} {Tool}},
	volume = {18},
	issn = {0965-0393},
	url = {https://doi.org/10.1088/0965-0393/18/1/015012},
	doi = {10.1088/0965-0393/18/1/015012},
	language = {en},
	number = {1},
	urldate = {2026-01-27},
	journal = {Modelling Simul. Mater. Sci. Eng.},
	author = {Stukowski, Alexander},
	month = dec,
	year = {2009},
	pages = {015012},
}

@article{maras_global_2016,
	title = {Global transition path search for dislocation formation in {Ge} on {Si}(001)},
	volume = {205},
	issn = {0010-4655},
	url = {https://www.sciencedirect.com/science/article/pii/S0010465516300893},
	doi = {10.1016/j.cpc.2016.04.001},
	urldate = {2026-01-27},
	journal = {Comput. Phys. Commun.},
	author = {Maras, E. and Trushin, O. and Stukowski, A. and Ala-Nissila, T. and Jónsson, H.},
	month = aug,
	year = {2016},
	pages = {13--21},
}

@article{obayashi_persistent_2022,
	title = {Persistent {Homology} {Analysis} for {Materials} {Research} and {Persistent} {Homology} {Software}: {HomCloud}},
	volume = {91},
	issn = {0031-9015},
	shorttitle = {Persistent {Homology} {Analysis} for {Materials} {Research} and {Persistent} {Homology} {Software}},
	url = {https://journals.jps.jp/doi/full/10.7566/JPSJ.91.091013},
	doi = {10.7566/JPSJ.91.091013},
	number = {9},
	urldate = {2026-01-27},
	journal = {J. Phys. Soc. Jpn.},
	publisher = {The Physical Society of Japan},
	author = {Obayashi, Ippei and Nakamura, Takenobu and Hiraoka, Yasuaki},
	month = sep,
	year = {2022},
	pages = {091013},
}

@article{fahy_theoretical_1987,
  title = {Theoretical total-energy study of the transformation of graphite into hexagonal diamond},
  author = {Fahy, S. and Louie, Steven G. and Cohen, Marvin L.},
  journal = {Phys. Rev. B},
  volume = {35},
  issue = {14},
  pages = {7623--7626},
  numpages = {0},
  year = {1987},
  month = {May},
  publisher = {American Physical Society},
  doi = {10.1103/PhysRevB.35.7623},
  url = {https://link.aps.org/doi/10.1103/PhysRevB.35.7623}
}

@article{scandalo_pressure_1995,
  title = {Pressure-Induced Transformation Path of Graphite to Diamond},
  author = {Scandolo, S. and Bernasconi, M. and Chiarotti, G. L. and Focher, P. and Tosatti, E.},
  journal = {Phys. Rev. Lett.},
  volume = {74},
  issue = {20},
  pages = {4015--4018},
  numpages = {0},
  year = {1995},
  month = {May},
  publisher = {American Physical Society},
  doi = {10.1103/PhysRevLett.74.4015},
  url = {https://link.aps.org/doi/10.1103/PhysRevLett.74.4015}
}

@article{khaliullin_nucleation_2011,
  title = {Nucleation mechanism for the direct graphite-to-diamond phase transition},
  author = {Khaliullin, R. Z. and Eshet, H. and Kühne, T. D. and Behler, J. and Parrinello, M.},
  journal = {Nat. Mater.},
  volume = {10},
  issue = {2011},
  pages = {693–-697},
  numpages = {0},
  year = {2011},
  month = {September},
  publisher = {Nature Springer},
  doi = {10.1038/nmat3078},
  url = {https://www.nature.com/articles/nmat3078#citeas}
}

@article{neuman_fractality,
  title = {Fractality and percolation in the phase transition of polycrystalline graphite},
  author = {Neuman, Vilmos and Malinowska-Trzmielak, Zuzanna and Wilson, Mark},
  journal = {Phys. Rev. B},
  pages = {},
  year = {2026},
  month = {Jul},
  publisher = {American Physical Society},
  doi = {10.1103/6zv5-drsn},
  url = {https://link.aps.org/doi/10.1103/6zv5-drsn}
}
\end{document}